\documentclass[12pt, preprint]{aastex}

\newcommand{\lunits}{$\rm erg~s^{-1}$}
\newcommand{\funits}{$\rm erg~cm^{-2}~s^{-1}$}
\newcommand{\cunits}{$\rm cm^{-2}$}
\newcommand{\chandra}{{\it Chandra~}}
\newcommand{\asca}{{\it ASCA~}}
\newcommand{\rosat}{{\it ROSAT~}}

\makeatletter

\newenvironment{inlinefigure}{%
\def\@captype{inlinefigure}%
\noindent\begin{minipage}{\linewidth}\begin{center}}
{\end{center}\end{minipage}\smallskip}
\makeatother

\shorttitle{Chandra Observations of IRAS00317-2142}
\shortauthors{Georgantopoulos et al.}

\begin{document}
\title{X-ray luminous galaxies I.
 {\it Chandra} Observations of IRAS00317-2142}

\author{I.~Georgantopoulos$\!$\altaffilmark{1,2} 
A.~Zezas$\!$\altaffilmark{3} 
M.~J.~Ward$\!$\altaffilmark{4} 
}

\altaffiltext{1}{Visiting Scientist, Harvard-Smithsonian Center for
Astrophysics, 60 Garden Street, Cambridge, MA02138}
\altaffiltext{2}{Institute of Astronomy \& Astrophysics, 
National Observatory of Athens, Palaia Penteli, 15236, Athens, Greece}
\altaffiltext{3}{Harvard-Smithsonian Center for Astrophysics, 60 
 Garden Street, Cambridge, MA02138}
\altaffiltext{4}{X-ray Astronomy Group, Department of Physics \&
 Astronomy, University of Leicester, LE1 7RH, UK}

\begin{abstract}
 We present {\it Chandra} observations of the enigmatic galaxy IRAS00317-2142, 
 which is classified as a star-forming galaxy on the basis 
 of the ionization level of its emission lines. 
 However, a weak broad $H\alpha$ wing and a high X-ray luminosity 
 give away the presence of an active nucleus. 
 The {\it Chandra} image reveals a nuclear point
 source ($\rm L_{(2-10 keV)} \sim 6\times10^{41} \rm erg~s^{-1}$), 
 contributing over 80 \% of the galaxy X-ray counts in the 
 0.3-8 keV band. This is surrounded by some fainter nebulosity
 extending up to 6 kpc. 
 The nucleus does not show evidence for  short-term variability. 
 However, we detect long term variations between the {\it ROSAT}, {\it ASCA} 
 and {\it Chandra} epoch. Indeed, 
 the source has decreased its flux by over a factor of  25 
 in a period of about  10 years.
 The nuclear X-ray spectrum is well represented by a 
  power-law with a photon index of 
 $\Gamma=1.91^{+0.17}_{-0.15}$ while the extended emission 
 by a Raymond-Smith component with a temperature of 
 $\rm \sim 0.6 keV$. 
  We find no evidence for the presence of an Fe line.  
 The nucleus is absorbed by an intrinsic column density 
 of $N_H\sim 8 \times 10^{20} \rm cm^{-2}$.  
 Thus the {\it Chandra} observations  
 suggest that at least the X-ray emission   
 is dominated by a type-1 AGN.
 Then the observed optical spectrum can be explained due to 
 the masking of the nucleus by the powerful star-forming 
 component. 
 These together with previous  X-ray observations of 
 galaxies with no clear signs of AGN activity in the optical,     
 (eg NGC6240) cast doubt on the 
 optical classification scheme and have implications 
 for the nature  of the 'normal' galaxies detected in deep 
 {\it Chandra} X-ray surveys.

\end{abstract}


\section{Introduction}
 Deep {\it ROSAT} (Boyle et al. 1995) and {\it Chandra} surveys 
 (Mushotzky et al. 2000, Brandt et al. 2000) have 
 discovered large numbers of apparently ``normal'' galaxies. 
 The optical spectra of these galaxies do not show evidence 
 for either broad or high excitation lines.  
 However, in many cases 
 the high X-ray luminosities observed ($L_x>10^{42}\rm erg~s^{-1}$) 
 suggest the possible presence of an AGN. 
 The nature of these galaxies remains unknown
 although they are generally believed to be associated 
 with an obscured active nucleus. 
 Unfortunately, no conclusive answers can be given 
 as these are faint and therefore there is a pressing 
 need to discover  their counterparts, if any,  in the nearby Universe.  
 It is often claimed (Barger et al. 2001) that these sources    
 may resemble the nearby galaxy NGC6240. 
 This galaxy offers one of the few  examples 
 in the nearby Universe where although the optical spectrum 
 (Veilleux et al. 1995) 
 gives an unambiguous LINER classification, the 
 X-ray spectrum reveals the presence of a heavily absorbed 
 AGN (Vignati et al. 1999). 
 Another class of nearby galaxies which has many  
 similarities to those 
 discovered in deep {\it ROSAT} and {\it Chandra} surveys  
 is the 'composite' 
 class of objects discovered by Veron et al. (1981)
 and Moran et al. (1996). 
 Their main characteristic is that their 
 [OIII] lines are broader than all other narrow lines 
 forbidden or permitted (Moran et al. 1996). 
 The diagnostic emission line diagrams
 of Veilleux \& Osterbrock (1987) classify
 these as star-forming galaxies. Yet, some of the composites
 present weak $H_\alpha$ broad wings and their X-ray luminosities 
 usually exceed $10^{42}$ $\rm erg~s^{-1}$ (Moran et al. 1996) 
 which is considered as the upper limit for the X-ray emission 
 of bona-fide star-forming galaxies (Halpern et al. 1995).    

 One of the most luminous ``composite'' objects 
 in the Moran et al. (1996) sample 
 is IRAS00317-2142 with a luminosity of $\sim 10^{43}$ \lunits 
 at a redshift of $z=0.0268$. This galaxy belongs to 
 the small group of galaxies HCG4. 
 The optical spectrum of IRAS00317-2142 has been 
 discussed in detail by Coziol et al. (1993) 
 and Moran et al. (1996).  
 The diagnostic emission line ratios
 ($H_\alpha/[NII]$ versus $H_\beta/[OIII]$)  
 classify it as a star-forming galaxy. 
 Nevertheless, its high X-ray luminosity 
 as well as a faint $H\alpha$ wing 
 would classify this as an AGN.
 Georgantopoulos (2000) presented {\it ASCA} 
 hard X-ray observations of this object. 
 The spectrum could be well fitted by a 
 single power-law  with $\Gamma\sim 1.8$
 while the ratio of the X-ray to the 
 broad $H_\alpha$ luminosity is 
 $\rm L_{(2-10 keV)}/L_{H\alpha}\sim2$.
 Both the above are typical of Seyfert nuclei
 and therefore suggest the presence of an
 unobscured AGN. 
 
 In this paper, we present {\it Chandra} ACIS-S observations 
 of IRAS00317-2142. The excellent spatial resolution of
 {\it Chandra} combined with its high effective area 
 and good energy resolution can be used  to shed more light on the 
 nature of this enigmatic AGN and elucidate  
 the origin of the disagreement between the X-ray and 
 the optical line ratio classification.

\section{Observations and Data Reduction}
IRAS00317-2142 was observed using 
 the Advanced CCD Imaging Spectrometer, ACIS-S,
  (Garmire et al. 2001)  
 onboard {\it Chandra} (Weisskopf et al. 1996). 
 The observation date 
 was 31-08-2001 (Sequence Number 700382). 
 We use the cleaned event file provided 
 by the standard pipeline processing. 
 The resulting exposure time is 19723 sec. 
 The target coordinates are (J2000)
 $\alpha = 00h34m13.8s$, $\delta =-21^\circ26m21s$.
 The whole galaxy falls on only one chip (S3).
  Charge Transfer Inefficiency (CTI) problems do not 
 affect our observations as S3 is a back-illuminated chip.  
 Each CCD chip subtends an 8.3 arcmin square on the sky
 while the pixel size is $0.5$ arcsec. 
 The spatial resolution on-axis is 0.5 arcsec FWHM. 
 The ACIS-S spectral resolution is 
 $\sim$100 eV (FWHM) at 1.5 keV.
 We observed in  $1/2$-subarray mode  
 in order to minimize pile-up problems.
 For our source's count rate the expected pile-up
 is only $\sim4\%$.  
 We used only Grade 0,2,3,4 and 6 events in the analysis.     
 Images, spectra an lightcurves have been created  
 using the {\sl CIAO v2.2} software.  
 The imaging and timing analysis  
 were performed using the {\sl SHERPA} 
 software. For the spectral fitting 
 we use the {\sl XSPEC v11} software package. 
 Throughout this paper, we adopt  
 a Hubble constant of $H_o= 65 \rm km~s^{-1}~Mpc^{-1}$.

\section{Analysis}

\subsection{Imaging}
 In Figure 1  we show the X-ray contours in the total band 
 (0.3-8 keV) overlayed on a Digital Sky Survey  image.
 The X-ray contours used are those from  the adaptively smoothed
 image created using the {\sl CIAO csmooth} routine.   
 Most of the  X-ray emission emanates from the 
 source coinciding with the optical nucleus. 
 The coordinates of the X-ray nuclear source are 
  $\alpha = 00h34m13.6s$, $\delta =-21^\circ26m18s$. 
 Its count rate is 0.054 $\rm cts~s^{-1}$ 
 in the 0.3-8 keV band,   
 corresponding to a flux of  $3.2\times10^{-13}$ \funits
 in the same band.
 The nucleus contributes the majority of the 
 galaxy counts in the 0.3-8 keV band within 5 arcsec (about 80 \%).
 The unobscured flux is $3.9\times10^{-13}$ \funits 
 corresponding to a luminosity of  $\rm L(0.3-8 keV)\approx
 7\times10^{41}$ \lunits. 
 For the conversion from count rate to flux 
 we used a a power-law spectrum with $\Gamma=1.9$
 and a column density of $N_H=8\times 10^{20}$ $\rm cm^{-2}$
 (see section 3.3).   

 An X-ray nebulosity can be clearly seen extending out to 
 at least 5 arcsec corresponding to 
 $\sim 6$ kpc. The extended emission within a circular
 region of 5 arcsec radius (excluding the nuclear emission i.e.
 a circular region 
 of 1.5 arcsec radius) has a count rate of 
 0.014 $\rm cts~s^{-1}$ in the 0.3-8 keV band.  
 The extended X-ray emission in the hard 2-8 keV band 
 is very small (0.001 $\rm cts~s^{-1}$ 
 in the above region).    
 The luminosity of the extended X-ray emission 
 is $\rm L(0.3-8 keV) \approx 7\times10^{40}$ \lunits, 
  assuming a Raymond-Smith spectrum 
 with a temperature of kT=0.3 keV.
 We use the {\sl SHERPA} software 
 to fit the radial profile of the extended emission. 
 We find that a Gaussian with a FWHM=3.5 arcsec 
 fits well the extended emission profile. 
    
 Finally, two point sources eastwards of the 
 nucleus are detected at distances of about 10 
 and 16 kpc respectively. Their coordinates are 
  $\alpha = 00h34m14.1s$, $\delta =-21^\circ26m24s$
 and  $\alpha = 00h34m14.4s$, $\delta =-21^\circ 26m26s$ (J2000).
 Their count rates are  
 $6.6\times 10^{-4}$ and $1.3\times 10^{-3}$ $\rm cts~s^{-1}$ (0.3-8 keV) 
 corresponding to luminosities of 
 $L_x\sim 6\times10^{39}$ \lunits and $L_x\sim 1\times 10^{40}$
 \lunits respectively. 
 These ultra-luminous X-ray sources are most likely 
 associated with X-ray binaries 
 (see Fabbiano, Murray \& Zezas 2001 and references therein).  

\begin{inlinefigure}
\label{image}
\epsscale{1}
\rotatebox{0}{\plotone{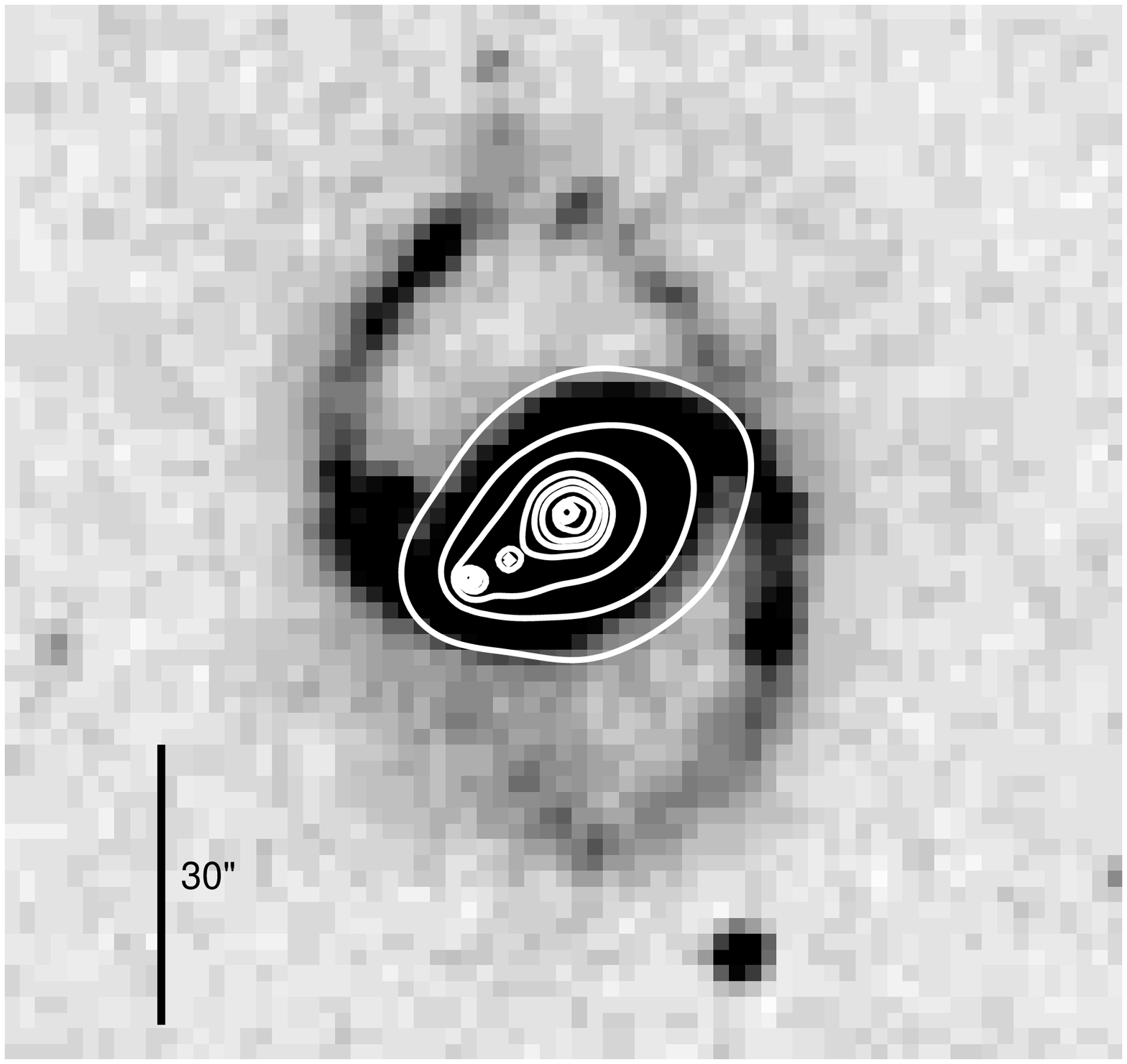}}
\figcaption{The X-ray contours from the 
 adaptively smoothed 0.3-8 keV 
 {\it Chandra} image overlayed on the 
 optical DSS image.}
 \end{inlinefigure}

\subsection{Timing}
 We first examine whether our source presents any  
 long term variability. 
 IRAS00317-2142 has been first observed in 1990 
 during  the {\it ROSAT} 
 all-sky survey (exposure 340 s). Subsequently, it 
 has been observed in a {\it ROSAT} PSPC pointed observation, 
 between June 22 and 23 1992, with an exposure time 
 of 9.3 ks. The 1-2 keV unabsorbed fluxes obtained 
 in these two {\it ROSAT} observations are  
 $2.0\pm0.46 \times10^{-12}$ 
 and $2.1\pm 0.05 \times 10^{-12}$ \funits 
 implying that the source exhibited no 
 variability in this two year period. 
 The 1-2 keV flux has decreased by a factor 
 of about 25 within a period of 10 years ie between 
 the {\it ROSAT} and the {\it Chandra} observation.

 {\it ASCA} observed IRAS00317-2142 between 
 December 11 and 12 1995. The 2-10 keV X-ray flux  
 was $8\times10^{-13}$ \funits.
 Hence, the 2-10 keV flux has decreased   
 by a factor of about 5  between the {\it ASCA} 
 and the {\it Chandra} epoch. 
 The long term light curve of IRAS00317-2142 
 is shown on Figure 2. We plot the 
 unabsorbed flux in the 1-2 keV energy range   
 as this is the common energy band    
 between {\it ROSAT}, {\it ASCA} and {\it Chandra}. 
 The conversion between count rate to flux 
 for all four points has been performed 
 assuming a single power-law spectrum 
 with a photon index $\Gamma=1.9$ absorbed by 
 a column density of $N_H=8\times10^{20}\rm cm^{-2}$
 (the best fit \chandra spectral model).

  Although the temporal sampling is sparse 
 the data are consistent with an AGN component that has almost 
 faded away by the time of the \chandra observation. 
 A similar behaviour has been observed in the 
 case of NGC4051 (Guainazzi et al. 1998)
 and NGC2992 (Gilli et al. 2000).  
 Note that the extraction radius  in the case of the  \chandra 
 data is much smaller that those used in the \rosat and \asca analysis 
 (1 and 2 arcmin radius respectively).   
 However, this cannot explain the observed decrease in flux 
 between the \asca and \chandra epoch. 
 Indeed, even when we use a 1 arcmin extraction radius we obtain 
 an unabsorbed flux of  only $1\times10^{-13}$ \funits (1-2 keV).  

 Next, we check for the presence of short term 
 variability in our {\it Chandra} data.  
 We extract the light curve using a 
 1 arcsec radius region. This 
 encompasses  $\sim$90\% of the photons from a point source 
 at an energy of 1.5 keV (on-axis).  
 We group our data into 1 ksec time bins.  
 The light curve is shown in Figure 3. 
 The errors correspond to the 68\% 
 confidence level. There is no evidence for 
 variability. Indeed, when we fit a constant 
 to the observed count rate  we obtain a 
 $\chi^2$ of 23.1 for 20  degrees of freedom (dof). 
 This implies that the no-variability hypothesis 
 cannot be rejected at a statistically significant level
 (less than 90\%). 

\begin{inlinefigure}
\epsscale{1.0}
\plotone{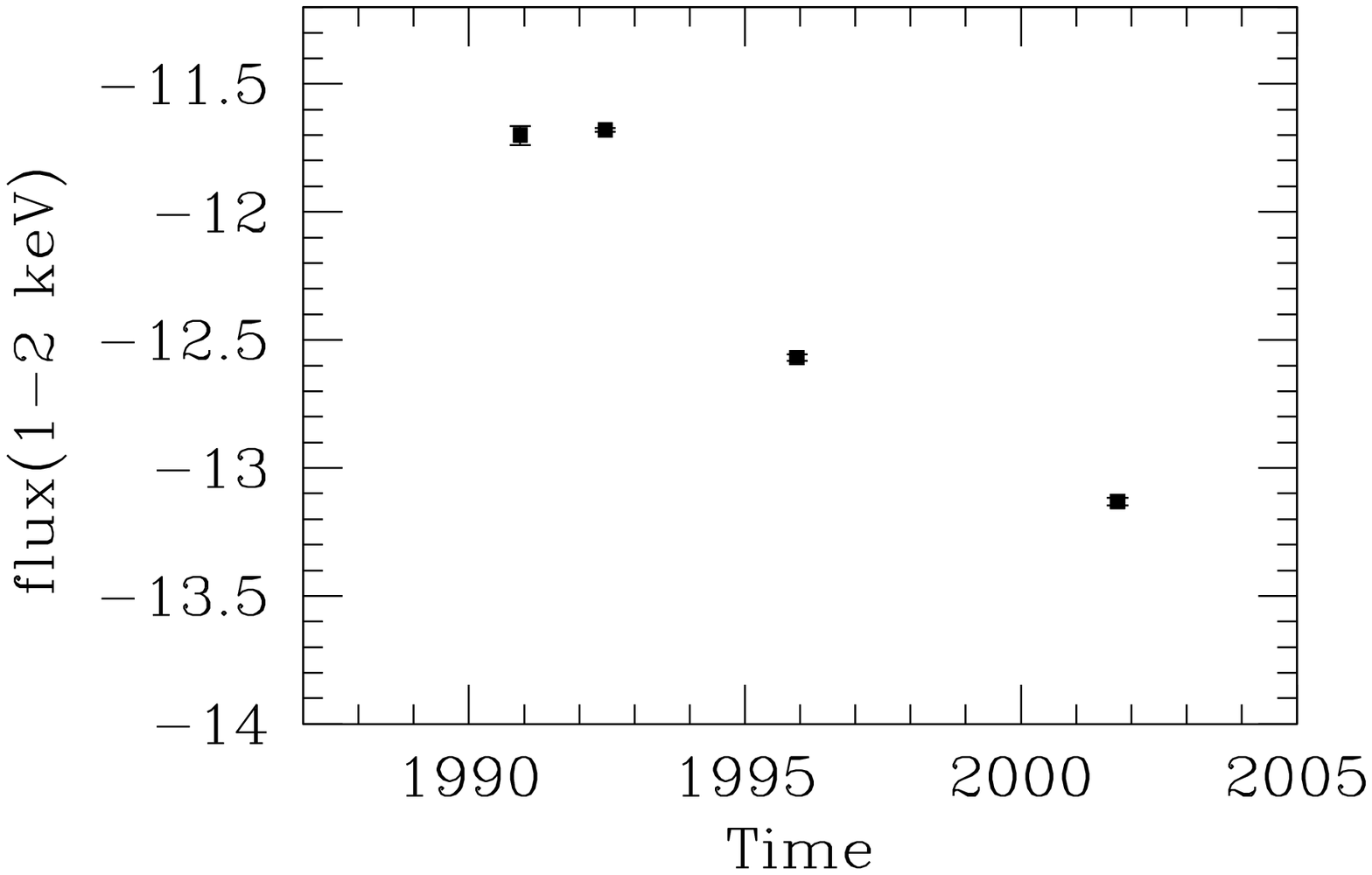}
\figcaption{The historical light curve 
 of the {\it unabsorbed} 1-2 keV flux. The points correspond 
 in chronological order to 
 a {\it ROSAT} (all-sky survey), {\it ROSAT} (pointed), 
{\it ASCA} and a {\it Chandra} observation}
\label{lc1} 
\end{inlinefigure}

\begin{inlinefigure}
\epsscale{1.1}
\rotatebox{0}{\plotone{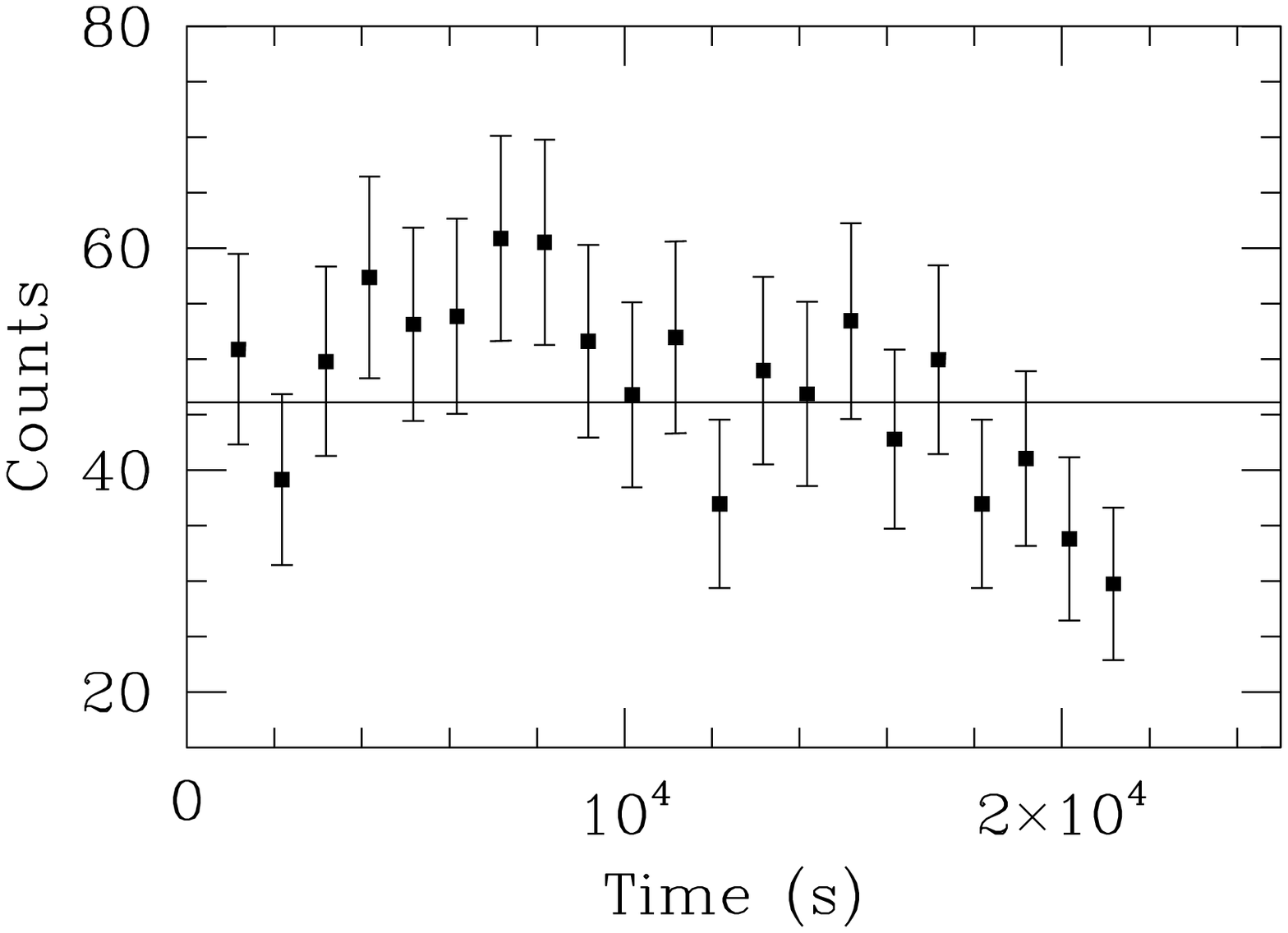}}
\figcaption{The short term light curve 
 grouped in 1 ks bins. The errors 
 correspond to $1\sigma$. The solid line 
 represents the best-fit average count rate.}
\label{lc2}
\end{inlinefigure}

\subsection{Spectral Fitting} 
We group the data so that there are at least 15 counts per energy bin. 
 The quoted errors to the best fitting spectral parameters 
 correspond to the 90 \% confidence level for one interesting 
 parameter.
 We discard data below 0.3 keV and above 10 keV
 due to both the low response and uncertainties in the 
 calibration matrix. 
 We present spectral fits for the following regions:

 \underline{a) the nuclear region.} Here, we  use a 1 arcsec region around the 
 central nuclear pixel. In order to take account of the 
   diffuse background originating from the galaxy itself, we use 
 background from 
 an annulus with an inner and outer radius of 2 and
 4 arcsec respectively. 
  We fit the data  using a single power-law model
 absorbed by two columns: one fixed to 
 the Galactic ($\rm N_H=1.5\times10^{20}$ Dickey \& Lockman 1990) 
 and one intrinsic which is left free to vary. 
 We find that this model provides an excellent fit ($\chi^2=47.3/49$)  
 to the data,
  see table 1 where all spectral fits results are listed.
 The power-law has the ``canonical''   slope, $\Gamma= 1.91^{+0.17}_{-0.15}$,
 while the intrinsic absorbing column is $8\pm3\times10^{20}\rm cm^{-2}$. 
 Unfortunately, there is not enough signal above 6 keV to provide strong 
 constraints on the Fe line. 
 Actually, the addition of a line at 6.4 keV (rest-frame) 
 results in no change in the $\chi^2$.
 The 90 \% upper limit on the equivalent width of 
 the Fe line is $\sim$1.4 keV. 
% We have also tried a warm absorber model ({\small ABSORI} in XSPEC).
% The $\chi^2$ has been reduced to 43.7/46 for 3 additional parameters,  
% not yielding a statistically significant 
% improvement. The ionised and the neutral columns are $\approx 1.2\times10^{22}$ 
% and $\approx 1.2\times10^{21}$ \cunits respectively. The ionization parameter 
% is $\xi=54^{+75}_{-34}$. The temperature is practically unconstrained
% and it was fixed to $\rm T=10^{6}$ K. 
 
 We finally note that  as it is possible that the above selected background region 
 includes some residual light from the nuclear source,
 as a cross-check we have also fitted 
 the spectrum using source-free background regions
 outside the galaxy but  still on the S3 chip. 
 The background subtracted spectra 
 give identical fits as the point source 
 overwhelms both the diffuse emission and 
 any small amounts of scattered light from the wings of the PSF. 
 In particular, the intrinsic column density 
 is $N_H=8^{+3}_{-2}\times 10^{20} \rm cm^{-2}$ 
 while the power-law slope is 
 $\Gamma=1.94^{+0.17}_{-0.16}$ in excellent agreement 
 with the fits presented above.

\begin{table*}
\caption{The spectral fits 
\label{tab}}
\centering
\begin{tabular}{cccccc}
\tableline\tableline
      & $\Gamma$ & $N_H$ & kT  &  {\sl Z}& $\chi^2$/dof \\
      &    & $10^{22}\rm cm^{-2}$ & keV &         &              \\
\tableline
Nucleus&  $1.91^{+0.17}_{-0.15}$  & $0.08^{+0.03}_{-0.03}$ & - &- & 47.3/49  \\
Diffuse &  -   & $0.15^{+0.20}_{-0.08}$ & $0.59^{+0.18}_{-0.42}$ & $0.04^{+0.06}_{-0.03}$& 12.2/7  \\
Nucleus+Diffuse   & $2.01^{+0.10}_{-0.10}$ & $0.11^{+0.07}_{-0.02}$ & 
      $0.30^{+0.04}_{-0.04}$ & $1^{+1.5}_{-0.7}$ & 62.9/62   \\
\tableline
\end{tabular}
\end{table*}

 \underline{b) The ``diffuse'' emission.} Here we use an annulus 
 with inner and outer radius of 2 and 5 arcsec 
 respectively. Again we use source-free regions 
 for the background subtraction.  
 We fit a Raymond-Smith (RS) model to the data 
 absorbed by two column densities, the Galactic 
 and an intrinsic one which is let free to vary.
 The fit yields $\chi^2=12.2/7$. 
 Some residuals above 2 keV could be either due to a second 
 harder component from X-ray binaries or 
 from residual light from the nuclear point source.    
 We find $\rm kT=0.59^{+0.18}_{-0.42} keV$ 
 for the temperature of the thermal component 
 and $N_H=1.5^{+2}_{-0.8}\times 10^{21}$ $\rm cm^{-2}$.
 The abundance  is abnormally low $0.04^{+0.06}_{-0.03}$ 
 possibly suggesting the presence of a multi-temperature medium 
 (see eg Buote 2000).

 \underline{ c) Combined diffuse and nuclear emission}. 
 Although we have already analysed the emission 
 from the nucleus and the 
 diffuse emission separately, for the sake of completeness
   we treat here the total galaxy emission as well. 
 We use a source region of 5 
 arcsec radius which encompasses both the nuclear and the 
 extended emission.  Note however, that this extraction radius 
 does not include the two point sources (ULXs)  
 eatsward of the nucleus.  
 The background is taken from  source-free regions
 on the S3 chip. 
 We use a model consisting of  a 
 power-law and a thermal (RS) component.
 We assume that {\it both} are absorbed 
 by the Galactic column density 
 while the power-law component 
 is obscured by an additional intrinsic absorber. 
 In Figure 4 we plot the spectral data,  
  together with the best-fit model 
 and $\chi^2$ residuals.  
 We find $\rm kT=0.30^{+0.04}_{-0.04} keV$,
 $\Gamma=2.01^{+0.10}_{-0.10}$, 
 $N_H=0.11^{+0.07}_{-0.02} \rm cm^{-2}$
 and an abundance $1^{+1.5}_{-0.7}$.
 The fit is good yielding  $\chi^2=62.9/62$. 
  When we assume that both the power-law and the thermal 
 component are obscured by the intrinsic column density 
 we obtain a similar fit with $\chi^2=62.7/62$. 
 Therefore, on the basis of the spectral fitting alone 
 it is difficult to decide on the location of the 
 absorbing screen.

\begin{inlinefigure}
\epsscale{0.7}
\rotatebox{270}{\plotone{f4.eps}}
\figcaption{The X-ray spectrum for the 
 whole galaxy region  together with the 
 best-fit spectrum, ie a power-law plus RS absorbed 
 by the Galactic and an intrinsic column density. 
 The $\chi^2$ residuals are plotted as well in the lower panel}
\label{spectrum}
\end{inlinefigure}

\section{Discussion}

The {\it Chandra}  observations allowed to us to disentangle the 
 nuclear from the extended star-forming component
 in this peculiar AGN. 
It becomes obvious that the nucleus contributes
 the bulk of the X-ray emission,
 producing over 80\%   of the photon counts 
 in the total 0.3-8 keV band. 
 The nuclear spectrum is well fit by a single power-law
 with $\Gamma\approx 1.9$ absorbed by an intrinsic column density 
 of $N_H\approx 8\times 10^{20}$ \cunits. 
 Although one has to take 
 into account the use of different spectral models 
 and the degeneracy between the power-law 
 and the column density, this is somewhat higher than the 
 column observed  at the \rosat 1992 observation,
 $N_H\approx 2\times10^{20}$, (Georgantopoulos  2000).
 Interestingly, the observed column in the \chandra  
 observation is more consistent with the column derived 
 from the Balmer decrement of the narrow line 
 region, $N_H\sim 10^{21}$ \cunits (Georgantopoulos 2000), 
 suggesting that the absorbing screen is located further 
 away from the nucleus. 
  
 It is well known that star-forming regions co-exist with AGN
 (eg Levenson et al. 2001, Della Ceca et al. 2001, Pappa et al. 2002). 
 IRAS00317-2142 is probably the extreme case of this class 
 where the emission from star-forming processes almost completely masks the 
 active nucleus in the optical.  
 The optical spectrum shows a low degree of ionization.
 The ratio of the optical emission lines 
  ($H_\alpha/[NII]$ against $H_\beta/[OIII]$)    
 place IRAS00317-2142 in the  star-forming galaxy regime.  
 However, the detection of a weak broad emission component  
to the $H\alpha$ line suggests the presence of an AGN. 
 Note that this weak broad emission would hardly be detected 
 at lower fluxes. 
 Actually 3 out of the 7 composite galaxies in Moran et al. (1996)
 do not present evidence for a broad line.  
 We discuss below a few models for the 
 origin of the X-ray  and optical emission in these objects,
 starting from the most possible one.  

 \underline{A strong star-forming component} 
 Moran et al. (1996) favoured a scenario in which the 
 star-forming component  is masking the weak nucleus 
 in the optical while in the X-rays it is the 
 nuclear emission which dominates. 
 The \chandra observations have shown that 
 in the X-ray band this is clearly the case. 
 In the optical a strong star-forming component is 
 observed. The integrated galaxy light 
  has $M_B\sim -21$ and  very blue colours 
 with U-B=-0.6 (Coziol et al. 1993).
 This component is also detected in X-rays having 
 $L_x\sim 7\times 10^{40}$ \lunits. 
 The 2 arcsec slit used by Moran et al. (1996) 
 corresponds to a large area on the galaxy ($\sim 2.5 kpc$) 
 and thus the masking of a 'normal' AGN by the powerful 
 star-forming component offers the most 
 plausible scenario.   
  
 \underline{An obscured nucleus}.
 The subtle AGN features in the optical spectrum 
 could be due to a weak UV ionizing 
 radiation field. For example circumnuclear 
 obscuration could cause this effect. 
 However, the observed X-ray column 
 ($\sim10^{21}\rm cm^{-2}$) corresponds 
 to only a couple of magnitudes of absorption in the 
 U band. Therefore one would need a 
 more contrived model where the 
 optical is obscured while the X-rays are not
 (eg a warm absorber model).   
 According to this the X-ray emission emerges  
 relatively unattenuated through an ionised column density 
 while 'pockets' of dust in this absorber are responsible 
 for the partial extinction of the optical radiation. 
 Still, the X-ray spectrum reveals no absorption edges 
 due to ionized matter. Moreover, in the standard AGN unification 
 model, the Narrow Line Region,  
 above the torus should be ionised  by the 
 nuclear radiation and hence we would expect 
 our galaxy to be classified as a Seyfert in the 
 standard line ratio diagnostic diagrams. Only in a geometry where the  
 nucleus is {\it spherically} covered by the obscuring screen,
 we  should expect the suppression of high excitation lines.
  However, an argument against such a 
  scenario comes from the
 $\rm L_{(2-10 keV)}-L_{H\alpha}$ relation.  
 Ward et al. (1988) found that in the case 
 of unobscured AGN,  
 the above two luminosities are tightly correlated. 
 In our case, the 
 2-10 keV X-ray luminosity observed by \asca (December 1995) 
 and the optical broad $H\alpha$ luminosity (obtained one year earlier) 
 follow the Ward et al. (1988) relation (Georgantopoulos 2000). 
 This implies that the optical radiation is not heavily attenuated, 
 provided of course that IRAS00317-2142 follows the 
 same spectral energy distribution as the 
 AGN in the sample studied by Ward et al. (1988).   
 Again the fact that the optical and the X-ray observations were 
 not simultaneous introduces some uncertainty 
 on the actual value of the $L_x/L_{H\alpha}$ ratio.  

\underline{X-ray outbursts}
 Large amplitude X-ray outbursts have been reported 
 in a few galaxies in the last few years. 
 For example the optically quiescent 
 galaxies IC3599 and NGC5905 have exhibited 
 outbursts of a factor of 100 (see Komossa \& Bade 1999
 and references therein). 
 The peak luminosities exceeded $L_X=10^{42}$ \lunits 
 while the spectra were very soft ($\Gamma\sim 4$). 
 These outbursts were interpreted as 
 tidal disruptions of stars by 
 supermassive black holes.
 NGC5905 and our object bear some similarities, namely   
 the absence of high excitation lines and the 
 very large variability amplitude. 
 On the other hand the ``X-ray outburst galaxies'' such 
 as NGC5905 show very rapid decay in their flux 
 (Komossa \& Bade 1999) while IRAS00317-2142 has 
 remained constant in flux between the two 
 \rosat observations. 
 Furthermore, our X-ray spectrum  
 is not as steep as these of Komossa \& Bade (1999) 
 but is more like that of a typical 
 AGN spectrum, probably ruling out such a model
 for the composite objects. 
 
 Regardless of the nature of the optical emission in composite objects,
 it is important to understand why the 
 optical spectrum only subtly indicates the 
 presence of the X-ray components. 
 This is probably not uncommon in the Low Luminosity AGN 
 regime. 
 The ``composite'' galaxies represent only one example 
 of this. Other even more striking examples are those 
 of the passive galaxies detected in deep \chandra and {\it XMM}  
 surveys (Mushotzky et al. 2000, Comastri et al. 2002).
 Most of these objects present no optical emission lines 
 while their X-ray emission rivals that of AGN. 
 Finally, the most well known case of the 
 optical/X-ray controversy classification   
 is probably the Compton-thick Seyfert-2 galaxy NGC6240 
 which appears to have a LINER optical spectrum.  
 All the above examples cast severe doubt on the 
 use of optical spectroscopy alone,
 in obtaining the identifications 
 of the sources in deep surveys.

\section{Summary} 
 We have presented \chandra observations of the 
 enigmatic galaxy  IRAS00317-2142.
 This object belongs to the 'composite' 
 class of \rosat all-sky survey sources studied by
 Moran et al. (1996). These galaxies have optical spectra 
 dominated by star-forming galaxy features while   
 in contrast their powerful X-ray emission  
 suggests the presence of an AGN.
 In this respect the composite class of galaxies 
 has many similarities to the numerous NLXGs  
 detected in deep \rosat and \chandra surveys.
 \chandra helped to shed more light on the 
 X-ray emission mechanisms in these objects.  
 The \chandra observations spatially resolve  
 the nuclear emission from the 
 surrounding extended component. 
 They show  that the X-ray emission 
 of IRAS00317-2142 is dominated by 
  a nuclear point source having a luminosity of 
  $\sim 6\times 10^{41}$ \lunits. 
  The luminosity of this object has rapidly declined
  (by  a factor of about 25) 
  in a period of about 10 years since the first \rosat 
 observation. This definitively argues for the 
 presence of an AGN in this object.  
  The nuclear component is absorbed by a 
 column density of $8\times 10^{20}$ \cunits 
 a factor of a few higher than the Galactic column.
 Thus the X-ray spectrum is suggestive of a type-1 AGN.   
 Extended  X-ray emission has been observed 
 around the nuclear source up to scales of 
 about 6 kpc contributing 
 less than 20 \% of the total counts. 
 The spectrum of this diffuse component  
 is  soft (described 
 by a RS model with a temperature of kT$\sim$0.6 keV)
 consistent with the diffuse gas temperatures encountered 
 in nearby star-forming galaxies (eg Read, Ponman \& Strickland 1997). 
 
 It remains then unclear why the optical spectrum 
 only subtly gives away the AGN features while 
 the X-ray emission is clearly dominated by the active nucleus.
 There are several models which can account for such a behaviour.
 The most plausible scenario is that the star-forming component 
 is very powerful and masks the AGN component in the optical.     
% Another, less likely, possibility is that the nucleus 
% has a low level of UV emission and hence 
% the broad lines are weak. 
% A more contrived scenario  would involve 
% a spherical coverage of 
% the nucleus by a warm absorber so that the X-ray emission is 
% not heavily attenuated.
 Our observations have implications for the 
 identifications of galaxies detected in deep \rosat and \chandra surveys. 
 It appears that at least some fraction of the 
 galaxies with apparently 'normal' star-forming galaxy optical spectra 
 can harbour an obscured nucleus.

\acknowledgements
We are grateful to the referee for many comments 
 and corrections. 
This work has been supported by the NASA grant   G01-2120X.

%%%%%%%%%%%%%%%%%%%%%%%%%%%%%%%%%%%%%%%%%%%%%%%%%%%%%%%%%%%%%%%%%%%%%%%%%%%%%

\end{document}